\documentclass[12pt]{iopart}

\usepackage{graphicx}
\usepackage{graphicx}

\begin{document}

\title{Temperature chaos and quenched heterogeneities}
 
\author{Paolo Barucca, Giorgio Parisi and Tommaso Rizzo}

\address{Dipartimento di Fisica Universit\'a di Roma ``La Sapienza"
Piazzale Aldo Moro 2, 00185, Roma, Italy
tel. +393285782858}
\ead{baruccap@gmail.com}
\begin{abstract}
We present a treatable generalization of the SK model which introduces
correlations in the elements of the coupling matrix through multiplicative
disorder on the single variables and investigate the consequences on the phase diagram. We define a generalized $q_{EA}$ parameter and test the structural stability of the SK results in this
correlated case evaluating the AT-line of the model. As main result we
demonstrate the increase of temperature chaos effects due to heterogeneities.

\end{abstract}

\section{Introduction}

Over the last few decades mean-field models of spin glasses have brought new concepts into physics that have been fundamental for the comprehension of a large number of complex phenomena belonging both to traditional physics and to interdisciplinary fields and many long-standing topics have benefited from the theoretical treatment: optimization problems, artificial and real neural networks, species dynamics, protein folding, glassy systems, disordered ferromagnets, jamming transition, opinion dynamics and so on
\cite{MPV, amit1985storing,mezard2009information,nishimori2001statistical,binder1986spin}. \\ 
In this paper we introduce a new spin glass model which is a solvable generalization of the Sherrington-Kirkpatrick (SK) model. We do this by multiplying each spin variable by a quenched weight representing heterogeneity. Many systems may present heterogeneous elements: particles systems with molecules of different sizes, assets with distinct intrinsic variability, heterogeneous modular networks, multi-atom alloys, lattice variables in a temperature field and so on. Nevertheless in most cases considering homogenous systems may suffice to capture most of the relevant features. On the other hand one may ask what is the difference between the heterogeneous and the homogeneous case. In particular one may enquire whether the solutions of a given model are structurally stable with respect to a transformation that changes single-variable parameters. \\
Here we focus on the SK model and we observe the consequences of the transformation we propose on temperature chaos \cite{jonason1998memory,bray1987chaotic}. The reason for such a study is that in recent years it has been hypothesized that local heterogeneity could be a basic mechanism for the enhancement of chaos in temperature. It was shown through the analysis of the SK model on diluted graphs \cite{parisi2010chaos, PhysRevLett.90.137201} that chaos is generally increased with respect to the fully-connected case with the remarkable exception of the Bethe lattice with bimodal interaction. The same exception holds also in large-deviation theory for the free energy \cite{parisi2010large, parisi2008large}. Analyzing sample-to-sample fluctuations of the free energy, the SK model and the model on the Bethe lattice with bimodal interaction share the same $N$-dependence, i.e. $N^{5/6}$, while in all other diluted cases with random couplings the fluctuations have been shown to be Gaussian.
The similarities in temperature chaos and in free-energy fluctuations of these two different mean-field models have been related to their local homogeneity. From this derives the hypothesis that introducing quenched single-spin heterogeneities in the SK model should have the same effects as dilution, i.e. chaos enhancement.
\\
Since we are considering a model with two sources of different quenched disorder we will first explain its physical interpretation then, through a detailed stability analysis, we will construct its phase diagram. Finally the two-temperature approach will be used to prove the main result. In  final remarks we will suggest further applications of the heterogeneity transformation in other spin glass models. \\

\section{Heterogeneity Transformation}

In the following we will consider a general mean-field spin glass (SG) model in the canonical ensemble and we will solve it with the replica method, comparing the homogeneous and the heterogeneous case. Formalizing the introductory description, the heterogeneity transformation (HT) reads: $ S_{i}\rightarrow w_{i} S_{i}$. The Hamiltonian of a set of $N$ spin $ S_i$ interacting on a graph $G $ is then:
\begin{equation}
H=\sum\limits_{(i,j)\in E}w_{i}w_{j}J_{ij} S_{i} S_{j}
\end{equation}
where $J_{ij}$ can be any symmetric matrix of interaction, deterministic or sampled from a distribution. $E$ is the set of the edges and $w_i$'s are $N$ i.i.d. weights.\\
Writing the Hamiltonian in terms of local fields we find: 
\begin{equation*} 
H=\frac{1}{2}\sum\limits_{i}^N w_{i} S_{i}h_{i} 
\end{equation*}    
 where the $h_{i}$ are defined by, $h_{i}=\sum\limits_{j|(i,j)\in E}J_{ij}w_{j} S_{j}$ 
 If we consider detailed balance for a single spin flip dynamics at finite temperature we can see: 
 \medskip{}
 \begin{equation*}  
  \frac{P( S_{i}\rightarrow- S_{i})}{P(- S_{i}\rightarrow S_{i})}=\frac{\pi(- S_{i},\{ S\}_{i})}{\pi( S_{i},\{ S\}_{i})}=e^{2\beta w_{i}h_{i}}
 \end{equation*} 
\medskip{}
where $P( S_{i}\rightarrow- S_{i})$ is the flip probability for site $i$ and $\pi(- S_{i},\{ S\}_{i})$ is the Gibbs measure for the flipped configuration. 
Considering that $w_{i}$ is quenched in each sample and that the distribution of $h_i$ is the same for each spin the quantity $\beta w_{i}$ has an effect similar to a local temperature localized on the site $i$. A given field $h$ has a stronger effect on spin with larger $w_i$ with respect to those with smaller weights.
\\
Furthermore in order to recover site dilution, which is closely related to the HT we may consider a distribution of weights $	\tilde{\rho}(w)$ such that:
\begin{equation*}   
\tilde{\rho}(w)=p\delta(w)+(1-p)\rho(w)
\end{equation*}    
In general the HT doesn't turn discrete spins (hard spins) into continuous ones (soft spins). If we suppose a continuous spectrum for the weights acting on hard spins each of them will still assume two values, $\{-1,+1\}$. The reason is simple: we are dealing with quenched disorder, while spins remain annealed variables. From the form of the transformation it is easy to see its applicability to any spin system (i.e. p-spin, Heisenberg model, Potts model).

\section{Weighted SK model (WSK)}

In this section we will solve the weighted SK model defined by the interaction Hamiltonian:

\begin{equation*}  
H=\sum\limits_{i,j}^Nw_{i}w_{j}J_{ij}S_{i}S_{j}
\end{equation*} 
where $J_{ij}$ is a matrix of i.i.d gaussian elements with mean $\frac{J_0}{N}$
and variance $\overline{J_{ij}^{2}}=\frac{J}{N}$ whereas $w_{i}$ are i.i.d variables with mean equal to one and finite variance
$\overline{w_{i}^{2}}=\frac{1+d^2}{2}$.
\\
With standard replica techniques we can
derive an integral expression for the free energy density $F$ to which we can apply the
saddle-point method.

\begin{equation*}   
\overline{Z^{n}}=\overline{\Tr \left\{ \exp\left(\beta\sum\limits_{i<j}^NJ_{ij}w_{i}w_{j}\sum\limits_a^n S_{i}^{a} S_{i}^{a}\right) \right\}}=\exp(-NF(q_{ab}^{*},m_a^{*}))
\end{equation*} 
where $q_{ab}^{*}$ and $m_{a}^{*}$ are the values at the saddle-point of the observables $q_{ab} = \frac{1}{N}\sum\limits_{i}^N S_{i}^{a} S_{i}^{b}$ and $m_{a} = \frac{1}{N}\sum\limits_{i}^N S_{i}^{a}$. 
In this case we find $F(q,m)$ to be:

\begin{equation*}  
F(q_{ab},m_{a})=\frac{\beta^{2}}{2}\sum\limits_{a<b}^nq_{ab}^{2}+\frac{\beta}{2}\sum\limits_{a}^nm_{a}^{2}-\log(L))
\end{equation*}    
where $L$ is the partition function in the configuration space of replicas:

\begin{equation}  
L=\Tr\left\{\int Dw\exp\left[\beta^2 J^{2}\sum\limits_{a<b}^nq_{ab}w^{2}S^{a}S^{b}+\beta J_0\sum\limits_{a}^nwm_aS^{a}\right]\right\}\label{partL}
\end{equation}    
we use $Dx$ to indicate the measure of integration of the variable $x$.
From the stationarity equations we find:
\begin{eqnarray}   
q_{ab}&=&J^2\langle w^{2}S^{a}S^{b}\rangle_{L}  \label{eqQea}\\
m_{a}&=&J_0\langle w \,S^{a}\rangle_{L}\nonumber 
\end{eqnarray}    
where $\langle \rangle_{L}$ denotes the average with respect to the distribution defined by the integrand of $L$ (\ref{partL}).
A physical interpretation for $q$ is as the Edward-Anderson order parameter, $q_{EA}=\frac{1}{N}\sum\limits_{i=1}^N\overline{ \langle S_{i}\rangle^{2}}$, our order parameter (\ref{eqQea}) can be associated to a reweighted form,
$\tilde{q}_{EA}=\frac{1}{N}\sum\limits_{i=1}^N\overline{w_{i}^{2}\langle S_{i}\rangle^{2}}$ instead.
\subsection{Replica-symmetric calculation}

We now perform a replica-symmetric (RS) analysis of the saddle-point equations to locate the transition temperature and to show at this level the differences with the SK equations. Therefore we equate all the off-diagonal terms of the overlap matrix and all the magnetizations of different replica $m_a$: 
\begin{eqnarray*}  
q_{ab}&=&q(1-\delta_{ab})\\
m_{a}&=&m
\end{eqnarray*}   
From this ansatz we are able to compute the saddle-point expression as a function of two variables, $F[m,q]$: 
\begin{eqnarray*}  
F(q,m)&=&-\frac{\beta^{2}J^{2}}{4}(\overline{w^{2}}-q)^{2}+\frac{\beta J_{0}m^{2}}{2}+\\
&&-\int Dw\int Dz\log(\cosh(\beta w(J\sqrt{q}z+J_{0}m+h))-\log2
\end{eqnarray*} 
where $Dz=\frac{e^{-\frac{z^2}{2}}}{\sqrt{2\pi}}dz$ is the Gaussian measure and the magnetic field $h$ has been introduced to show the equations in their full generality.\\
Finally we set the derivatives to zero and get two coupled equations for $m$ and $q$ that could have been obtained also with cavity arguments:
\begin{equation} 
q=\int Dw\int Dzw^{2}\tanh^{2}(\beta w(J\sqrt{q}z+J_{0}m+h))
\label{RSq} \end{equation}
\begin{equation}  
m=\int Dw\int Dzw\tanh(\beta w(J\sqrt{q}z+J_{0}m+h))
\end{equation}    
In order to simplify the analysis of the phase diagram we focus on the zero field and zero mean coupling case, $J_{0}=0$ and $h=0$.
Consequently the equation for $m$ always admits one $m=0$ stable solution and we only have to deal with the self-consistent equation
for $q$. \\
Only the trivial solution $q=0$ exists for high temperature, a non-zero solution appears below $T_{c}=\sqrt{\overline{w^{4}}}$, which is a function of the ensemble of weights through parameter $d$.
\\
From the RS analysis of the saddle point equations it is possible to show that in the spin glass phase $\tilde{q}_{EA}-\overline{w^2}q_{EA}>0$. Hence a correlation appears in the unweighted overlap of the same site between two replicas and its weight, i.e. sites with bigger weights have bigger overlaps. This can be explained noticing that weights act as local effective temperatures therefore larger spins act as if they were at lower $T$. 
\subsection{AT Line and phase diagram}
Following the de Almeida-Thouless approach  \cite{dAT} we can perform the spectral analysis of the Hessian
of the free energy at the replica symmetric solution [A1].
\\
The similarities between the WSK and the SK are such that the eigenvalue
structure of the Hessian is conserved, in particular the replicon
eigenvalue still remains the only one that turns negative below
the critical temperature in the RS phase. Using AT notation the replicon
is found to be: 
\begin{equation*}   
\lambda=P-2Q+R
\end{equation*}    
where 

\begin{equation*}   
P=1-(\beta J)^{2}(\overline{w^{4}}-\overline{\langle w^{2} S^{a} S^{b}\rangle}^{2})
\end{equation*}    
\begin{equation*}  
Q=-(\beta J)^{2}(\overline{\langle w^{4} S^{b} S^{c}\rangle}-\overline{\langle w^{2} S^{a} S^{b}\rangle}^{2})
\end{equation*}    
\begin{equation*}  
R=-(\beta J)^{2}(\overline{\langle w^{4} S^{a} S^{b} S^{c} S^{d}\rangle}-\overline{\langle w^{2} S^{a} S^{b}\rangle}^{2})
\end{equation*}    
with some algebra we can express the $\lambda>0$, in a compact form:
 
\begin{equation}   
(\beta J)^{-2}>\int Dw\int Dzw^{4}(1-\tanh^{2}(\beta w(J\sqrt{q}z+J_{0}m+h)))^2\label{dATcond}
\end{equation}    
Hence we find that there exists a critical temperature $T_{AT}$ at
which the replica symmetric stationary point is no longer stable.
In the SK model, with $J_{0}=0$, the RS critical temperature,
$T_{c}$, and the de Almeida-Thouless critical temperature coincide.
\\
Analyzing the AT-equation and comparing with the RS equation (\ref{RSq}) it is possible to show that similarly
in WSK model the positivity condition (\ref{dATcond}) is always violated by second order terms in $q$ below $T_c$. The replicon eigenvalue is always negative below $T_c$.
\begin{figure}[h!]
  \centering
      \includegraphics[scale = .5]{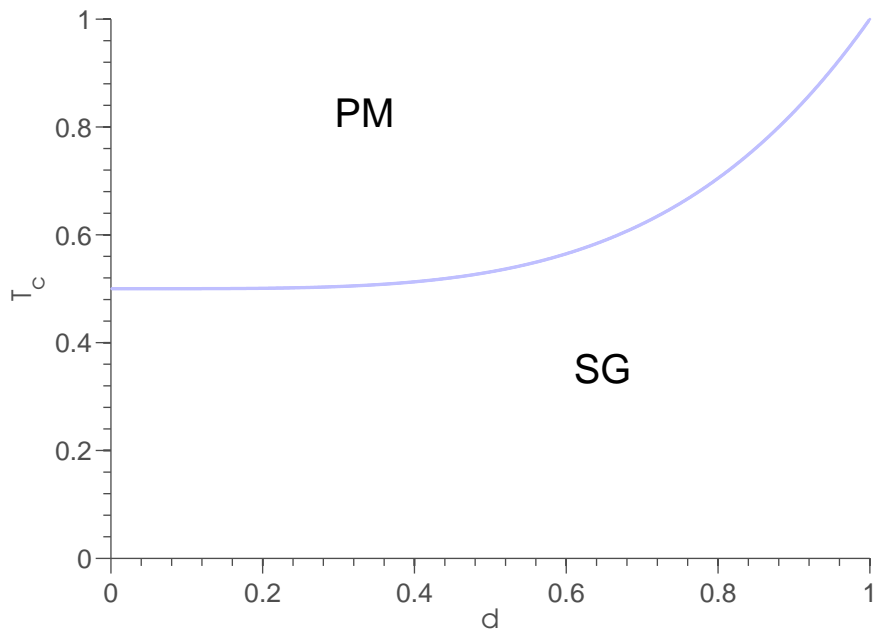}
  \caption{Critical line in the case of a bimodal distribution of equiprobable weights $\{1,d\}$. In the case $d=0$ we end up with a SK model with $T_c$ equal to $0.5$ since half of spins are missing.}
\end{figure}
The stability analysis has revealed the need for replica-symmetry breaking (RSB). The form of the multiplicative disorder allows us to compute also the 1RSB expression for the saddle-point free energy that reads:

\begin{eqnarray*}   
\beta f_{1RSB}&=&\frac{\beta^{2}J^{2}}{4}((m_{1}-1)q_{1}^{2}-m_{1}q_{0}^{2}+2\overline{w^{2}}q_{1}-(\overline{w^{2}})^{2})+\frac{\beta J_{0}m^{2}}{2}-\log(2)+\\
&&+\int Dw\int Dz(\frac{1}{m_{1}}\log(\int Dv\cosh^{m_{1}}(\beta w\mathcal{H}[z]))
\end{eqnarray*}    
where $\mathcal{H}[z]=(J\sqrt{q_{0}}z+J\sqrt{q_{1}-q_{0}}v+J_{0}m+h)$ and where $Dv$ is the Gaussian measure.
The form of $f_{1RSB}$ shows how the HT can be easily taken into account in RSB calculations allowing to control its effects at each step of the computation.
Below the AT-line we expect symmetry to be broken infinitely many times, like in the SK model. So in order to fully characterize the low temperature phase of the WSK model we have to compute the fRSB expression for the free energy.\\
Following standard RSB steps it is possible to verify:

\begin{eqnarray*}  
\beta f_{fRSB}[q(x)]&=&-\frac{\beta^{2}J^{2}}{4}\{(\overline{w^{2}})^{2}+\int_{0}^{1}dxq^{2}(x)-2\overline{w^{2}}q(1)\}+\\
&&-\int Dw\int Dzf_{w}(0,\sqrt{q(0)}wz)
\end{eqnarray*}    
with $f_{w}(x,h)$ satisfying the Parisi equation:
\begin{equation*}   
\frac{\partial f_{w}(x,h)}{\partial x}=-\frac{J^{2}w^{2}}{2}\frac{dq}{dx}\{\frac{\partial^{2}f_{w}}{\partial x^{2}}+x(\frac{\partial f_{w}}{\partial h})^{2}\}
\end{equation*}    
with boundary condition:
\begin{equation*}   
f_w(1,h)=\log{2}\cosh{h}
\end{equation*}    
This set of equations is quite similar to the fRSB SK solution \cite{PhysRevLett.43.1754} yet more difficult to
solve numerically since for a generic distribution of the $w$'s we
have to solve the Parisi differential equation and plug the solution
into the free energy expression and average with the measure $Dw$. 
\section{Free energy expansion near $T_{C}=J\sqrt{\overline{w^{4}}}$}
The replica calculation has explicitly shown that WSK model is indeed a generalized SK model, in the sense that its free energy can be expressed by the extremization of a functional of the overlap matrix $Q_{ab}$. \\
Since we are dealing with a second-order phase transition we expect that chaos effects, in which we are interested, and also other statical and dynamical properties \cite{PhysRevLett.108.085702} depend on the free energy expansion for small $Q_{ab}$, i.e. near $T_c$. In particular we are mainly interested on the coefficients of such an expansion.  
The functional can indeed be written in the form:

\begin{eqnarray}  
F[Q]&=&-\frac{\tau}{2}\Tr{Q^{2}}-\frac{\omega}{6}\Tr{Q^{3}}+\nonumber\\
       &&-\frac{u}{12}\sum\limits_{a,b}^nQ_{ab}^{4}-\frac{v}{8}\Tr{Q^{4}}+\frac{y}{4}\sum\limits_{a,b,c}^nQ_{ab}^{2}Q_{ac}^{2} \label{exps}
\end{eqnarray} 
where $\tau=\frac{1-T^2/T_c^2}{2}$.\\
In the WSK case starting from the replica expression:

\begin{eqnarray*}   
\overline{Z^{n}}&=&A\int dq_{ab}\exp\left[-\frac{N}{2}\beta^{2}\sum\limits_{a<b}^nq_{ab}^{2}+\right.\\
&&\left.+N\log\left(\Tr\left\{\int Dw\exp\left(\beta^{2}\sum\limits_{a<b}^nq_{ab}w^{2}S^{a}S^{b}\right)\right\}\right)\right]
\end{eqnarray*}    
Expanding near the critical temperature $T_{c}$ we introduce $Q_{ab}=\beta^{2}q_{ab}$ and obtain in terms of the $p(w)$ distribution's momenta $\overline{w^{k}}$ the following expressions for the coefficients [A2]:

\begin{equation*}   
\omega=\frac{\overline{w^{6}}}{\overline{w^{4}}}
\end{equation*}    

\begin{equation*}   
u=v=y=\frac{\overline{w^{8}}}{\overline{w^{4}}}
\end{equation*}    
In generalized SK models the relevant non trivial solution  $q(x)$ \cite{parisi1980order,crisanti2002analysis} can be written in terms of the expansion coefficients \cite{rizzo2001against} leading to:
\begin{eqnarray*}  
q(x)=\frac{\overline{w^6}}{\overline{w^8}}\frac{x}{2} \qquad (0 \le x \le x_1=2q(1))\\
q(x)=q(1)\qquad (x_1 \le x \le 1)
\end{eqnarray*}   
\\
where $q(1)=\theta \frac{\overline{w^4}}{\overline{w^6}}$ and $\theta=\frac{T^2-T^2_c}{2T^2_c}$.

\begin{figure}[h!]
  \centering
      \includegraphics[scale = .5]{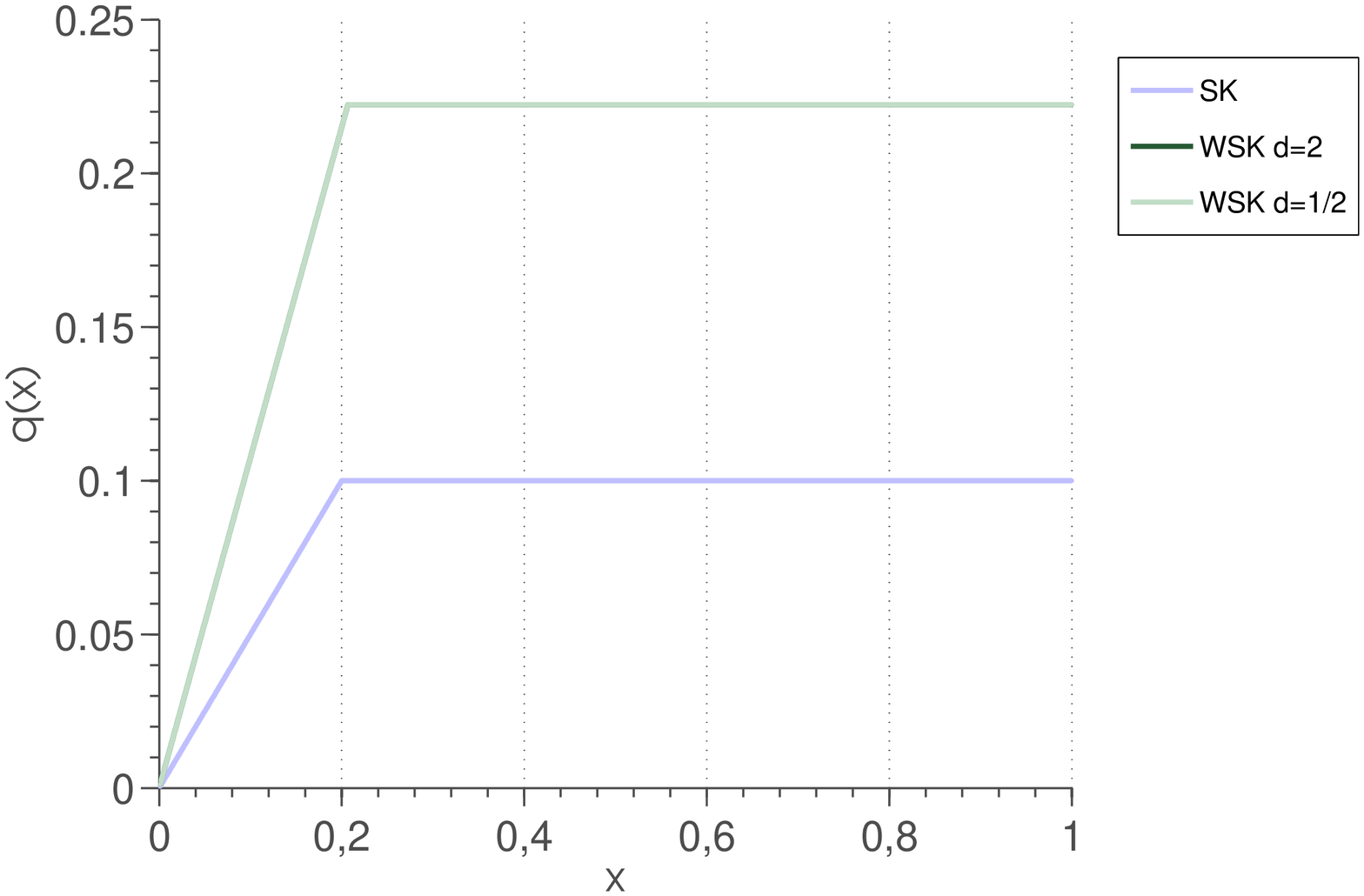}
  \caption{Relevant behavior of q(x) near $T_c$. The curves for $d=2$ and $d=1/2$ coincide due to the symmetry of the model [A2]}
\end{figure}

\section{Temperature Chaos}
Chaos in temperature in the SG literature is the vanishing of correlations between equilibrium states at different temperatures in the thermodynamical limit \cite{krzakala2002chaotic,kondor1989chaos,bray1987chaotic}.
In this chapter we study temperature chaos and through the two-temperature approach we show its enhancement as already proved in the link diluted case, thus confirming expectations: a strict relation between local heterogeneity and chaos in temperature in disordered systems.

\subsection{The two-temperature approach}
RSB theory states that at any temperature below the critical one there are infinitely many pure states organized in an ultrametric space. The correlation between two states is measured by the overlap $q_{ab}$ according to the Parisi function $P(q)$  which can take values between zero and $q_{EA}$ which is the state self-overlap.
On the other hand chaos in temperature concerns a generalization of $P(q)$ , the function $P_J^{\beta_1 \beta_2}(q)$, which quantifies overlaps among states at different temperature. It reads:

\begin{equation*}   
P_J^{\beta_1 \beta_2}(q)= \frac{\sum\limits_{\{S^{(1)}\}\{ S^{(2)}\}}\delta( q-\frac{1}{N}\sum\limits_i  S^{(1)}_i S^{(2)}_i)\exp(-\beta_1 H_J\{ S^{(1)}\} -\beta_2 H_J\{S^{(2)}\})}{\sum\limits_{\{\tau\}\{ S\}}\exp[-\beta_1 H_J\{ S^{(1)}\} -\beta_2 H_J\{S^{(2)}\}]}
\end{equation*}  
There is chaos if $P_J^{\beta_1 \beta_2}(q)$ equals a Dirac delta in zero in the thermodynamical limit. 
Given the formula above we can quantitatively verify the presence of chaos in a disordered model, from the large-deviations relation:
\begin{equation*}
P^{\beta_1 \beta_2}(q) \sim \exp{(-N\Delta F_{12})}
\end{equation*}
where we consider the free energy shift $\Delta F_{12}=F_{12}-F_1-F_2$ which is given by the difference between the constrained free energy functional $F_{12}(q,\beta_1,\beta_2)$ and the unconstrained free energies of the two coupled systems averaged over the disorder.
The constrained free energy reads:

\begin{equation*}   
F_{12}(q,\beta_1,\beta_2)=-{1 \over N} \overline{ \ln \sum\limits_{\{S^{(1)}\}\{ S^{(2)}\}}\delta\left(N\,q-\sum\limits_i  S^{(1)}_i S^{(2)}_i\right)\exp(-\beta_1 H_J\{ S^{(1)}\} -\beta_2 H_J\{S^{(2)}\})}
\end{equation*}
In the generalized SK model it was proved that in the small $q$ and small differences in rescaled temperatures $\Delta\tau=|\tau_1-\tau_2|$ limit the leading corrections in the free energy shift read \cite{rizzo2001against, rizzo2009chaos, parisi2010large}:\\
if $ q>>|\tau_1 - \tau_2|$\\

\begin{equation}   
\Delta F_{12}(q)= A\, \beta_c^6{|q|^3}(\tau_1-\tau_2)^2 \ \ \ A={ u \over 6 \omega }\left({v \over \omega^2}-c_{12}\right) \ , 
\label{DFq3}
\end{equation}    
if $ q<<|\tau_1 - \tau_2|$\\
\begin{equation}   
\Delta F_{12}(q)=B\, \beta_c^4q^2|\tau_1-\tau_2|^3\ \ \ 
B={u^{1/2} \over   \omega 2^{3/2} \pi}\left({v \over \omega^2}-c_{12}\right)^{3/2}
\label{DFq2} 
\end{equation}    
where $c_{12}$ is defined by the expansion$\tau_{12}=\frac{1}{2}(\tau_1+\tau_2)+\frac{c_{12}}{4}(\tau_1-\tau_2)^2 + O(\tau^3)$, in which $\tau_{12}=\frac{1-T_1T_2/T_c^2}{2}$. \\
In the SK model chaos effects are weak in finite-size systems because the quantity $\left({v \over \omega^2}-c_{12}\right)$ vanishes thus higher order corrections have been computed \cite{PhysRevLett.90.137201} leading to the expression:

\begin{equation*}   
\Delta F_{12}(q) ={12 \over 35}\ |q|^7\Delta T^2
\end{equation*}    
\subsection{Chaos in WSK}
Starting from the approach we used for the single-temperature free energy expansion we can consider the free energy for a system composed of two WSK subsystem with different
temperatures and constrained weighted overlap $q$:

\begin{eqnarray*}   
F_{12}[\hat{Q}]&=&\frac{1}{2}\overline{w^{4}}\{\tau_{1}Tr\hat{Q}_{1}^{2}+\tau_{2}Tr\hat{Q}_{2}^{2}+2\tau_{12}TrP^{2}+\frac{1}{3}\frac{\overline{w^{6}}}{\overline{w^{4}}}\Tr\hat{Q}^{3}+\\
&&+\frac{1}{6}\frac{\overline{w^{8}}}{\overline{w^{4}}}\sum\limits_{a,b}^n\hat{Q}_{ab}^{4}+\frac{1}{4}\frac{\overline{w^{8}}}{\overline{w^{4}}}\Tr\hat{Q}^{4}-\frac{1}{2}\frac{\overline{w^{8}}}{\overline{w^{4}}}\sum\limits_{a,b,c}^n\hat{Q}_{ab}^{2}\hat{Q}_{ac}^{2}\}
\end{eqnarray*}    
with $\tau_{1}=\frac{1-T_{1}^{2}/T_{c}^{2}}{2}$,$\tau_{2}=\frac{1-T_{2}^{2}/T_{c}^{2}}{2}$,$\tau_{12}=\frac{1-T_{1}T_{2}/T_{c}^{2}}{2}$,
$c_{12}=1$, remembering $T_{c}^{2}=\overline{w^{4}}$
\\

\begin{figure}[h!]
  \centering
      \includegraphics[scale = .4]{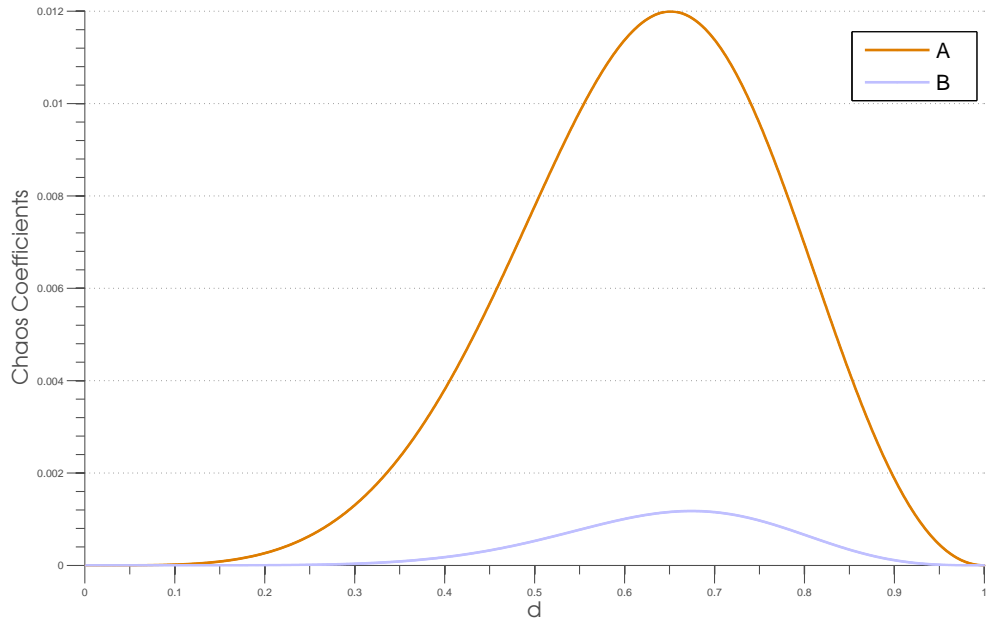}
  \caption{Coefficients curves for chaos in the two limits considered for a bimodal distribution in $\{1,d\}$. The curves have rather sharp maxima indicating the optimal values for chaos enhancement. The extremal values are zero since the SK limit is recovered.}
\end{figure}
In order to compare these results with the free energy shift formulas (\ref{DFq3})-(\ref{DFq2}) we have to express $A$ and $B$ in absolute units, thus we consider the rescaled overlap $\tilde{q}=q/\overline{w^2}$. Hence we end up with the following rescaled chaos parameters  $A$ and $B$:\\
if $ q>>|\tau_1 - \tau_2|$\\
\begin{equation*}   
\Delta F_{12}(\tilde{q})= A {\tilde{q}^3}(\tau_1-\tau_2)^2 \ \ A=\frac{(\overline{w^2})^3}{(\overline{w^4})^2}\frac{\overline{w^{8}}}{6\overline{w^{6}}}\left(\frac{\overline{w^{8}}\,\overline{w^{4}}}{(\overline{w^{6}})^{2}}-1\right)
\end{equation*}
whereas if $ q<<|\tau_1 - \tau_2|$\\
\begin{equation*}   
\Delta F_{12}(\tilde{q})=B{\tilde{q}^2}|\tau_1-\tau_2|^3\ \ \ 
B=\frac{(\overline{w^2})^2}{\sqrt{\overline{w^4}}}\frac{\sqrt{\overline{w^{8}}}}{2\sqrt{2}\pi\overline{w^{6}}}\left(\frac{\overline{w^{8}}\,\overline{w^{4}}}{(\overline{w^{6}})^{2}}-1\right)^{3/2} 
\end{equation*}    

From the Cauchy inequality, given a kernel distribution defining a
scalar product $\langle\rangle=\int dw[\cdot]p(w)$:

\begin{equation*}   
\langle f(w)g(w)\rangle^{2}<\langle f(w)^{2}\rangle\langle g(w)^{2}\rangle
\end{equation*}    
Taking $f(w)=w^{2}$ and $g(w)=w^{4}$, we notice that $A,B>0$, for any distribution different from a Dirac delta, thus demonstrating in the replica framework the chaos enhancement by local heterogeneities.

\section{Conclusions}
We introduced a generalized SK model by multiplying each spin of the system by a positive quenched variable $w_i$. The factorizable structure of the multiplicative heterogeneity transformation allowed us to use standard replica techniques to establish the phase diagram and to prove that it undergoes RSB below $T_c$. Thus we showed the structural stability of the eigenvalue behavior of the RS Hessian at any temperature. Once the similarity with SK model was displayed we demonstrated how this simple transformation is able to modify statical properties such as temperature chaos through the free energy expansion. As a consequence we enforced the general hypothesis of a connection between local heterogeneity due to random couplings, local topology or spin dependent features and temperature chaos. Conversely this supports the hypothesis that local homogeneity could be the reason for the similarities between SK and Bethe lattice model with bimodal interactions.  
Observing chaos in SK has been a long-lasting problem \cite{billoire2002overlap, aspelmeier2002temperature, billoire1999evidences} and only recently hints for the presence of such a phenomenon in finite-dimensional systems was revealed in Edward-Anderson model simulations \cite{fernandez2013temperature} through rare events analysis. The heterogeneity transformation we proposed is simple to implement and can be performed on any topology. Its effect on statical properties, such as temperature chaos, can be maximized through the choice of the distribution and its parameters. Heterogeneity in the WSK enhances by four orders of magnitude the free energy shift. Therefore in order to have low probabilities for large deviations in $q$, considering that $ P^{\beta_1 \beta_2}(q) \sim \exp{(-N\Delta F_{12})}$, $N$ must be four orders of magnitude than in the SK model. Hence substantially diminishing size requirements for simulations up to a linear prefactor given by the value of the coefficients $A$ and $B$.\\
In this work on the WSK model we have seen that theoretical calculation are not substantially modified by the HT yet statical properties of the model can vary. Therefore extending the transformation also to other spin-glass models could be a straightforward way to check their structural stability and test the existence of new physics, i.e. we are currently investigating the weighted version of the XOR-SAT random optimization problem in the UNSAT phase. 

\section*{Appendix}
\subsection*{A1. RS Stability analysis}
The RS saddle-point expansion (quadratic part):

\begin{equation*}
\Delta=\sum\limits_{(ab)(cd)}(\delta_{(ab)(cd)}-\beta^{2}(\overline{\langle w^{4}S^{a}S^{b}S^{c}S^{d}\rangle}-\overline{\langle w^{2}S^{a}S^{b}\rangle}\overline{\langle w^{2}S^{c}S^{d}\rangle}))\eta^{ab}\eta^{cd}
\end{equation*}
Using standard notation from the seminal paper \cite{dAT}:
\begin{equation*}
G^{(ab)(ab)}=P=1-\beta^{2}(\overline{\langle w^{4}\rangle}-\overline{\langle w^{2}S^{a}S^{b}\rangle}^{2})
\end{equation*}

\begin{equation*}
G^{(ab)(ac)}=Q=-\beta^{2}(\overline{\langle w^{4}S^{b}S^{c}\rangle}-\overline{\langle w^{2}S^{a}S^{b}\rangle}^{2})
\end{equation*}

\begin{equation*}
G^{(ab)(cd)}=R=-\beta^{2}(\overline{\langle w^{4}S^{a}S^{b}S^{c}S^{d}\rangle}-\overline{\langle w^{2}S^{a}S^{b}\rangle}^{2})
\end{equation*}

The replicon eigenvalue is:
\begin{equation*}
\lambda_{R}=P-2Q+R=1-\beta^{2}(\overline{w^{4}}-2\overline{\langle w^{4}q\rangle}+r)>0
\end{equation*}

\begin{equation*}
(\beta J)^{-2}>\int Dw\int Dzw^{4}(1-\tanh^{2}(\beta w(J\sqrt{q}z+J_{0}m+h)))^{2}
\end{equation*}

Transverse eigenvalues are:
\begin{equation*}
\lambda_{T}=\frac{1-P+4Q-3R\pm|P-4Q+3R|}{2}
\end{equation*}

\begin{eqnarray*}
P-4Q+3R &=&1-\beta^{2}(\overline{w^{4}}-4\overline{\langle w^{4}q\rangle}+3r) \\
&=&1-\beta^{2}\int Dw\int Dzw^{4}(1-\tanh^{2}(\beta w(J\sqrt{q}z+J_{0}m+h)))\\
&&\qquad \qquad \qquad \qquad \qquad (1-3\tanh^{2}(\beta w(J\sqrt{q}z+J_{0}m+h)))
\end{eqnarray*}
Numerically evaluating the equations we find that both sets of eigenvalues have the same temperature dependence as the SK model.

\subsection*{A2. Symmetry in temperature and disorder}
In this appendix we exhibit a transformation of the parameter which leaves the WSK free energy invariant and we show that the explicit expansion we computed is indeed symmetric, providing a good check for the result.
The free energy for the replicated system with constrained overlap matrix $q_{ab}$ reads:
\begin{equation*}   
F(q,\beta, p(w))=-{1 \over N} \overline{ \ln \sum\limits_{\{ S\}}\delta\left(N\,q_{ab}-\sum\limits_i w_i^2 S_i^a  S_i^b\right)\exp[-\beta\sum\limits_{(i,j)\in E}w_{i}w_{j}J_{ij}\sum\limits_{a} S_{i}^a S_{j}^a]}
\end{equation*}
where we explicitly indicated the dependence on the weight distribution $p(w)$. From the multiplicative form of the disorder we can consider the following change of parameters:
\begin{eqnarray*}
\beta &\rightarrow& \beta \,c\\ 
p(w) &\rightarrow& p(w\,\sqrt{c})\\ 
q &\rightarrow& q/c 
\end{eqnarray*}
where $c$ is arbitrary positive number. It is easy to verify that:
\begin{equation*}   
F(q,\beta, p(w))=F(q/c,\beta\,c, p(w\,\sqrt{c}))
\end{equation*}
So this equivalence must hold in full generality. In particular we check the obtained $F[q]$ expansion (\ref{exps}) and we find, a part from the first prefactor which doesn't contain weights and it's invariant, all the others are of the form $a_nw^{2n}\beta^{2n}q^n$. Performing the transformation we can see that:
\begin{equation*}
a_n(w/\sqrt{c})^{2n}(\beta\,c)^{2n}(q/c)^n=a_nw^{2n}\beta^{2n}q^n
\end{equation*}
Thus the symmetry holds at each order of the expansion and consequently it holds for all other thermodynamical quantities that can be derived from it. 

\section*{References}
\bibliographystyle{plain}
\bibliography{tcaqh}
\end{document}